\documentclass{article}

\usepackage{thumbpdf,lmodern,amsmath,amsthm, amssymb, float,listings}
\usepackage{graphicx}
\usepackage{tikz}
\usetikzlibrary{decorations.fractals}

\usepackage[authoryear]{natbib} 

\usepackage{framed}

\def\mm#1{\ensuremath{\boldsymbol{#1}}} 

\author{
  Janet van Niekerk, Haakon Bakka, H{\aa}vard Rue
  and Olaf Schenk\\KAUST and USI
}

\title{New frontiers in Bayesian modeling using the {INLA} package in {R}}

\begin{document}
\maketitle
\begin{abstract}
 The {INLA} package provides a tool for computationally
  efficient Bayesian modeling and inference for various widely used
  models, more formally the class of latent Gaussian models. It is a
  non-sampling based framework which provides approximate results for
  Bayesian inference, using sparse matrices. The swift uptake of this
  framework for Bayesian modeling is rooted in the computational
  efficiency of the approach and catalyzed by the demand presented by
  the big data era. In this paper, we present new developments within
  the {INLA} package with the aim to provide a computationally
  efficient mechanism for the Bayesian inference of relevant
  challenging situations. 
  \end{abstract}





\section[Introduction to the R-INLA project]{Introduction to the
  {R}-{INLA} project} \label{sec:intro} 

The {R-INLA} project is an evolving platform that hosts
various projects, all interlinked with respect to the {INLA}
package in {R}. This package is based on the INLA methodology
developed by \citet{rue2009}. This development revolutionized the
availability and applicability of Bayesian modeling approaches, even in
high dimensions, to practitioners and statisticians alike. The INLA
methodology ensures computational efficiency by using sparse
representations of high dimensional matrices used in latent Gaussian
models (LGMs). The computational efficiency of the method offers great
appeal to different fields of science and for various applications. In
ecology, \citet{quintero2018} studied bird diversity by using
{R-INLA} while \citet{braga2018} investigated environmental
relationships by incorporating phylogenetic information.
\citet{dalongeville2018} used R-INLA to genes specific to salinity in
the field of genomics (also see ). Air pollution was assessed with the
purpose of disease assessment by \citet{shaddick2018} while
\citet{rodriguez2018} used R-INLA to determine forest species
distributions. A study into fire occurrences was conducted by
\citet{podschwit2018} to develop a forecasting system with the use of
R-INLA. The effect of coral bleaching in the Great Barrier Reef on the
marine ecosystem was investigated by \citet{stuart2018}. In social
studies, R-INLA has been applied to study the state of education
\citep{graetz2018} and child growth \citep{osgood2018} in Africa.
These aforementioned works are but a few of many recent applications
of R-INLA. The pertinence of {R-INLA} is clear. We believe
that the new developments presented here will enable more applications
in an even broader context.

We present a brief conceptual framework of the INLA methodology.
Latent Gaussian models is a specific subset of
hierarchical Bayesian additive models. This class comprises of
well-known models such as mixed models, temporal and spatial models.
An LGM is defined as a model having a specific hierarchical structure,
as follows: The likelihood is conditionally independent based on the
likelihood parameters (hyper parameters), $\pmb{\theta}$ and the linear
predictors, $\eta_i$, such that the complete likelihood can be
expressed as
\begin{equation*}
\pi(\pmb{y}|\pmb{\eta},\pmb{\theta})=\prod_{i=1}^{N}
\pi(y_i|\eta_i(\pmb{\mathcal{X}}),\pmb{\theta}).
\end{equation*}
The linear predictor
is formulated as follows:
\begin{equation}
\eta_i=\beta_0+\pmb{\beta}^\top\pmb{X}_i+\pmb{u}_i(\pmb{z}_i)
\label{additive predictor}
\end{equation}
where $\pmb{\beta}$ represent the linear fixed effects of the
covariates $X$, $\pmb{\epsilon}$ is the unstructured random effects
and the unknown
non-linear functions $\pmb{u}$ of the covariates $\pmb{z}$ are the structured random effects. These include spatial effects, temporal effects, non-separable
spatio-temporal effects, frailties, subject or group-specific
intercepts and slopes etc. This class of models include most models
used in practice since time series models, spline models and spatial
models, amongst others, are all included within this class. The main
assumption is that the data, $\pmb{Y}$ is conditionally independent
given the partially observed latent field, $\pmb{\mathcal{X}}$ and
some hyper parameters $\pmb{\theta}_1$. The latent field
$\pmb{\mathcal{X}}$ is formed from the structured predictor as
$(\pmb{\beta},\pmb{u},\pmb{\eta})$ which forms a Gaussian Markov
random field with sparse precision matrix $\pmb{Q}(\pmb{\theta}_2)$,
i.e.\ $\pmb{\mathcal{X}}\sim N(\pmb{0},\pmb{Q}^{-1}(\pmb{\theta}_2))$.
A prior, $\pmb{\pi}(\pmb{\theta})$ can then be formulated for the set
of hyper parameters $\pmb{\theta}=(\pmb{\theta}_1,\pmb{\theta}_2)$. The
joint posterior distribution is then given by:
\begin{equation}
  \pmb{\pi}(\pmb{\mathcal{X}},\pmb{\theta})\varpropto\pmb{\pi}(\pmb{\theta})\pmb{\pi}
  (\pmb{\mathcal{X}}|\pmb{\theta})\prod_{i}\pi(Y_i|\pmb{\mathcal{X}},\pmb{\theta})
  \label{postINLA}
\end{equation}
The goal is to approximate the joint posterior density
(\ref{postINLA}) and subsequently compute the marginal posterior
densities, $\pmb{\pi}(\mathcal{X}_i|\pmb{Y}),i=1...n$ and
$\pmb{\pi}(\pmb{\theta}|\pmb{Y})$. Due to the possibility of a
non-Gaussian likelihood, the Laplace approximation is used to
approximate this analytically intractable joint posterior density. The
sparseness assumption on the precision matrix which characterizes the
latent Gaussian field ensures efficient computation
\cite{rue2005}.

In this paper we present some new developments within the {INLA}
package in the fields of complex survival models, spatio-temporal
models and high performance computing. In Section \ref{sec:surv} we discuss the
implementation of complex survival models including joint
longitudinal-survival models, competing risks models and multi-state
models. Each of these could incorporate spline, spatial, temporal or
clustering elements to mention a few. We then present the new
extensions in the spatio-temporal domain, non-separable
space-time models. Finally, we discuss how the {INLA} package is
adapted to a high performance computing environment using the {PARDISO}
library.



\section[Complex survival models using the INLA package]
{Complex survival models using the {INLA} package}\label{sec:surv}

Survival models are used extensively in clinical studies where the
time to a certain event is of interest. The hazard function, the
instantaneous risk of experiencing the event, is most often of
interest to estimate. More importantly, the effects of covariates on
the hazard function is of interest for causal inference. Parametric
and nonparametric approaches have been proposed to model the hazard
function, most are available in the {INLA} package. In this
section, we focus on more complex survival models and will not discuss
standard survival models (see \citet{martino2011}). 

\subsection{Joint longitudinal-survival models} \label{sec:jointmodels} 

A joint model comprises of two different likelihoods and these likelihoods are joined by shared random
effects (see \cite{wulfsohn1997,hu2003,guo2004}). Extensions of linear joint models like spatial
random effects and non-linear trajectories are used in the context of joint models to address
certain practical challenges (see \cite{zhou2008,ratcliffe2004,andrinopoulou2018}). 
Each of these new joint models is still
a latent Gaussian model and thus no special implementation package is
needed for each one (for more details see \citet{van2019}). Most
longitudinal likelihoods and hazard function assumptions can be facilitated in
this framework, leaving no need to develop a new implementation for
each set of assumptions.
\\
\\
Within the realm of joint longitudinal-survival models, users have a choice of various computational approaches. The {joineR} library in {R} is widely used to fit joint models from a frequentist point of view whereas the {JMBayes} library facilitates Bayesian estimation of joint models. The {joineR} library can even accomodate competing events in the survival submodel. In terms of partially linear joint models the {JointModel} library was developed to fit non-linear covariate effects in the longitudinal submodel using B-splines with a sieve approximation. 
The {bamlss} library can also be used to fit a partially linear joint model using a Markov Chain Monte Carlo (MCMC) approach. We, however, aim to show that most joint models (also with non-linear covariate effects) can be fitted using the {INLA} library, also including discrete and non-Gaussian continuous joint models. This provides the user with one computational tool for the Bayesian inference of most joint models, since our approach provides support for non-linear covariate effects through continuously-indexed splines as well as discrete and continuous spatial random effects. 

\subsubsection{Joint models as LGMs}

In this section, we present relevant details of the joint model as an LGM
as defined in Section \ref{sec:intro}, full details are available in
\citet{van2019}. We first present details of the two sub models, and
then focus on the joint model in its entirety. Suppose the hazard rate
for individual $i,i=1,...,N^S$ at time $s$ is defined by
$$
h_i(s)=h_0(s)\exp(\eta^S_i(s))
$$
where $h_0(s)$ is the baseline hazard function which can be
parametrically or non-parametrically specified and $\eta^S_i(s)$ is
the linear predictor, based on covariates, for individual $i$.
Currently, the exponential, weibull, log-Gaussian and log-Logistic survival distributions are
included in the {INLA} package, under the parametric
hazard function assumption. The Cox proportional hazards model is
included as a semi-parametric model resulting from a non parametric
constant baseline hazard in each of many time partitions (see \cite{cox1972}). In this case, the random
walk prior is adopted for the logarithm of the piece wise-constant
baseline hazard function, achieving a non parametric estimate of the baseline hazard function. Now
define
$$
f_i(s|\eta^S_i(s))=h_i(s)\exp\left(-\int_0^s h_i(u)du\right),
$$ 
then the likelihood for the survival sub model is
\begin{equation}
    \pi_S(\pmb s|\pmb{\eta}^S)=\prod_{i=1}^{N^S}\pi_i(s|\eta^S_i)
    =\prod_{i=1}^{N^S} f_i(s|\eta^S_i)^{c_i}[1-F_i(s|\eta^S_i)]^{1-c_i},
    \label{survlik}
\end{equation}
where $c_i=\mathcal{I}($non-censored observation$)$ indicates if an
observation is not censored. An observation is censored when the exact
event time is not observed but rather the most informative non-event
time. Right, left or interval censoring are common and can be
accommodated in our approach. The observations are thus a mixture of
event times and censored times, dependent on the status of each
individual.

Now, for the longitudinal data suppose that each individual has
$N_i, i=1,...,N^S$ observations $y_{ij},i=1,...,N^S,j=1,...,N_i$ for a
total longitudinal data set size of $N^L=\sum_{i=1}^{N^S}N_i$. We
specify the linear predictor $\eta^L_i(t)$, based on covariates at
time $t$, and a conditional density function $g(y_i|\eta^L_i(t))$ for
individual $i$ resulting in the likelihood for the longitudinal
sub model as
\begin{equation}
    \pi_L(\pmb{y}|\pmb{\eta}^L)=\prod_{l=1}^{N_L} g(y_l|\eta^L_l(t)).
    \label{longlik}
\end{equation}
Now consider the linear predictors of the joint model,
\begin{eqnarray}
  \eta^{L,J}_l(t)&=&\eta^L_l(t)\notag\\
  \eta^{S,J}_i(s)&=&\eta^S_i(s)+f(\eta^L_i(s)),
\label{jointmodel}
\end{eqnarray}
where $\eta^S$ and $\eta^L$ are of the form (\ref{additive predictor})
and $f:\Re\rightarrow\Re$ is a smooth function of $\eta^L_i(t)$. The
function $f$ facilitates the joint estimation of the models and can
assume various forms. A common approach is to use the entire
longitudinal linear predictor (see \citet{ibrahim2010}), while traditionally only
the subject-specific intercept and slope of the time effect have been
used i.e.\ $f(\eta^L_l(s))=\nu_1w_1+\nu_2w_2s$. In the latter we assume
the structure specified by \citet{henderson2000} as follows,
$$ 
\begin{bmatrix} 
    w_1 \\ w_2 
\end{bmatrix}\sim
N 
\begin{pmatrix} 
    \begin{bmatrix}
        0 \\ 0 
    \end{bmatrix}, 
    \begin{bmatrix}
        \sigma^2_{w_1} & \rho\sigma_{w_1}\sigma_{w_2}\\ 
        \rho\sigma_{w_1}\sigma_{w_2} &
        \sigma^2_{w_2} 
    \end{bmatrix} 
\end{pmatrix}.
$$
Note that either $\nu_1$ or $\nu_2$ can be defined to be zero if
desired.

Based on this reconstruction of the joint model, it was demonstrated
by \citet{van2019} that the joint model is indeed an LGM and can be
successfully applied with the {INLA} package.

To this end, we present two illustrative examples. Firstly, we use
data from a randomized clinical trial used to investigate the
efficacy of two antiretroviral drugs in HIV patients available in the
{JMBayes} package where $f(\eta^L_i(s))=\nu_1w_1+\nu_2w_2s$.
Secondly we present an example with a non-linear trajectory and
informative dropout event process with $f(\eta^L_i(s))=\nu\eta^L_i(s)$
from a prostate cancer study using post treatment PSA levels as a
longitudinal bio marker.

\textbf{Example 1. HIV antiretroviral treatments efficacy}\\
In this example the efficacy and safety of two antiretroviral treatments, Didanosine and Zalcitabine, is investigated and presented in \citet{guo2004}. This randomized trial includes only patients who had failed or were intolerant to Zidovudine (AZT) therapy. In the joint model, we use the same association structure as in \citet{guo2004}, i.e.\ 
\begin{eqnarray}
        \eta^{L,J}_l(t)&=&\eta^L_l(t)+w_1+w_2t\\
        \eta^{S,J}_i(s)&=&\eta^S_i(s)+\nu_1w_1+\nu_2w_2s.\label{jointex}
\end{eqnarray}
This model estimates the treatment effect on the survival as well as CD4 count jointly. We can then evaluate the treatments for efficacy in both endpoints by the inclusion thereof as a covariate in both sub models. The specific sub models are then 
\begin{eqnarray*}
        \eta^{L,J}_l(t)&=&\beta_0^L+\beta_1^L \text{Gender}+\beta_2^L \text{Drug}+\beta_3^L \text{Previous OI} +\beta_4^L \text{AZT Resistance} +w_1+w_2t\\
        \eta^{S,J}_i(s)&=&\beta_0^S+\beta_1^S \text{Gender}+\beta_2^S \text{Drug}+\beta_3^S \text{Previous OI} +\beta_4^S \text{AZT Resistance} +\nu_1w_1+\nu_2w_2s.
\end{eqnarray*}
The data is loaded and visualized by

\begin{lstlisting}
R> library("INLA")
R> inla.setOption(short.summary = TRUE)
R> data("aids", package = "JMbayes")
R> par(mfrow = c(1,2))
R> interaction.plot(aids$obstime[1:100],aids$patient[1:100],aids$CD4[1:100], 
+                   xlab="Time(years)", ylab="CD4 count",
+                   legend=F, col=c(1:467))
R> hist(aids$CD4,main="",xlab="CD4 count")
\end{lstlisting}
\begin{figure}[H]
	\centering
\includegraphics[width=12cm]{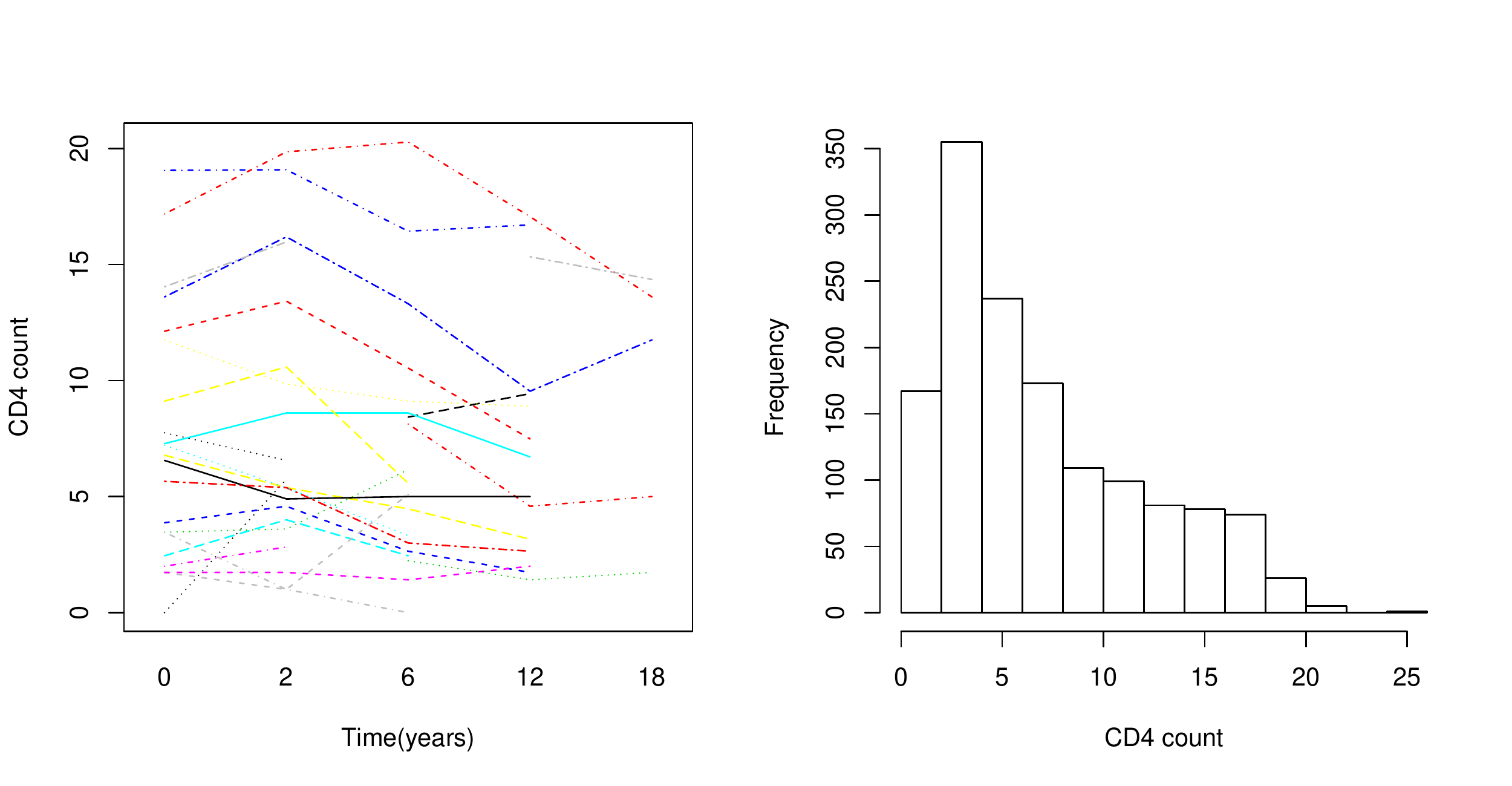}
\caption{Individual profiles and histogram of CD4 counts}
\label{CD4fig}
\end{figure}
In \citet{guo2004} the CD4 counts were transformed with the square root function to use the Gaussian distribution for the response model. In this example we use the original counts and assume a Poisson distribution instead. In Figure \ref{CD4fig} it is clear that no zero inflation is evident, although such phenomena could be incorporated into the model using a zero-inflated Poisson distribution for the longitudinal response (available as {zeroinflatedpoisson0} or {zeroinflatedpoisson1} for types $0$ and $1$, respectively). The individual CD4 trajectories are very different from one another and the need for individual-specific models are clear. This motivates the inclusion of subject-specific intercepts and slopes into the longitudinal sub model. In this example we assume the Weibull distribution for the survival times, although the exponential, log-Gaussian, log-Logistic or Cox proportional hazards assumptions could be used as well. We also rescale the time axis to the unit axis using the maximum time for this data set:

\begin{lstlisting}
R> data1 = aids
R> mtime = max(max(data1$Time),max(data1$obstime))
R> mtime

[1] 21.4
\end{lstlisting}
All the times hereafter should thus be rescaled to $(0;21.4)$ for interpretation. \\
Some preprocessing of the data is required to perform the joint analysis. The details are omitted here but the concept is illustrated in \eqref{jointdata}. Define,
\begin{eqnarray*}
\pmb{y}&=&\begin{bmatrix} y_1 & NA \\ y_2 & NA\\ ... & ...\\y_{N^L} & NA \\
NA & s_1\\ NA & s_2 \\ ...&...\\ NA  & s_{N}\end{bmatrix}\notag\\
\pmb{\beta}&=&[\beta_1^{L} \ \beta_1^{S} \ ...]\notag\\
\pmb{X}&=&\begin{bmatrix} x^L_{1,1} & 0 & ... \\ x^L_{1,2} & 0& ...  \\ ... & 0& ... \\x^L_{1,N^L} & 0 & ... \\
0 & x^S_{1,1}& ...  \\ 0 & x^S_{1,2}& ...  \\ 0 &...& ... \\0 & x^S_{1,N}& ...   \end{bmatrix}\notag\\
\end{eqnarray*}
\begin{eqnarray}
\pmb{u}(\pmb{t})&=&\begin{bmatrix} 
w_{1,1} & w_{2,1}t_1 & NA & NA\\ w_{1,1} & w_{2,1}t_2 & NA & NA \\  ... & ... & ... & ...\\ w_{1,N} & w_{2,N}t_{n^L} & NA & NA \\NA & NA & \nu_1w_{1,1} & \nu_2w_{2,1}s_1\\ NA & NA & \nu_1w_{1,2} & \nu_2w_{2,2}s_2\\... & ... & ... & ...\\NA & NA & \nu_1w_{1,N} & \nu_2w_{2,N}s_{N}  \end{bmatrix}.\label{jointdata}
\end{eqnarray}
Then the joint model in \eqref{jointex} is an LGM similar to \eqref{additive predictor}. In this paper, we use the pre-processed data available in the {INLA} package using the following code.
\begin{lstlisting}
R> joint.dataCD4 = readRDS(system.file("exampledata/cd4/jointdataCD4.rds",
+  package = "INLA"))
\end{lstlisting}
The joint model is fitted using the {inla} function with the defined {formula}. The {family} argument contains the information of the likelihood model(s) and subsequently the appropriate link function(s) for the linear predictor. Since the joint model consists of two likelihoods and hence two linear predictors, we specify the {poisson} distribution for the longitudinal series and the {weibull} distribution for the hazard rate.

\begin{lstlisting}
R> JointmodelCD4 = inla(Y ~ -1 + mu + l.gender + l.drug + l.prevOI +
+            l.AZT + s.gender + s.drug + s.prevOI + s.AZT + 
+            f(U11,model="iid2d",n=2*length(joint.dataCD4$mu)) +
+            f(U21,l.time,copy="U11",fixed=TRUE) + 
+            f(U12, copy="U11",fixed=FALSE) +
+            f(U22,s.time,copy="U11",fixed=FALSE), family = c("poisson",
+            "weibullsurv"), data = joint.dataCD4, verbose=FALSE,
+            control.compute = list(dic=TRUE))
R> summary(JointmodelCD4)


Fixed effects:
            mean    sd 0.025quant 0.5quant 0.975quant   mode kld
mu1        2.336 0.095      2.150    2.336      2.524  2.336   0
mu2       -1.171 0.296     -1.777   -1.162     -0.616 -1.145   0
l.gender2 -0.016 0.091     -0.195   -0.017      0.163 -0.017   0
l.drug2    0.059 0.053     -0.046    0.059      0.163  0.059   0
l.prevOI2 -0.692 0.067     -0.824   -0.692     -0.560 -0.691   0
l.AZT2    -0.020 0.070     -0.157   -0.020      0.117 -0.020   0
s.gender2 -0.340 0.247     -0.801   -0.348      0.169 -0.365   0
s.drug2    0.211 0.148     -0.078    0.211      0.502  0.211   0
s.prevOI2  1.286 0.228      0.849    1.282      1.745  1.274   0
s.AZT2     0.154 0.164     -0.166    0.153      0.479  0.151   0

Model hyperparameters:
                                     mean    sd 0.025quant 0.5quant
alpha parameter for weibullsurv[2]  1.279 0.056      1.169    1.279
Precision for U11 (component 1)     4.233 0.416      3.477    4.211
Precision for U11 (component 2)     3.811 2.517      0.777    3.239
Rho1:2 for U11                      0.095 0.305     -0.480    0.091
Beta for U12                       -1.048 0.218     -1.473   -1.049
Beta for U22                        0.977 0.281      0.418    0.981
                                   0.975quant   mode
alpha parameter for weibullsurv[2]      1.388  1.282
Precision for U11 (component 1)         5.111  4.168
Precision for U11 (component 2)        10.286  2.072
Rho1:2 for U11                          0.678  0.041
Beta for U12                           -0.618 -1.053
Beta for U22                            1.523  0.993

Deviance Information Criterion (DIC) ...............: 4703.56
Deviance Information Criterion (DIC, saturated) ....: 2149.36
Effective number of parameters .....................: 381.33
\end{lstlisting}

Similarly to \citet{guo2004}, the status of previous opportunistic infection ({prevOI}) is a significant covariate in both the longitudinal and survival models. The association between the longitudinal and survival models is significant. Entry into the study with a previous AIDS diagnosis results in lower CD4 counts and an increased hazard of death. There is a negative significant association between the initial value of the CD4 trajectory and the linear predictor of the survival model, which indicates a decreased hazard of death for individuals with higher CD4 counts at study entry. The positive association between the hazard rate and the slope of CD4 is an anomaly which is explained by the use of a Weibull survival model with an estimated shape parameter of $1.398$, which indicates an increase in hazard over time. The random time trend association we aim to capture in $\nu_2$ is thus construed with the shape parameter of the Weibull model.\\ \\
To use the model for patient-specific predictions we extract the necessary components from the latent field of the longitudinal and survival sub models. We use the data in {dataH} to calculate the survival functions and {dataL1} to illustrate the observed and estimated longitudinal trajectories.

\begin{lstlisting}
R> data1$Time = data1$Time/mtime
R> data1$obstime = data1$obstime/mtime
R> datas = data1[data1$obstime==0,]
R> datal = data1[,c(1,4:12)]
R> ns = nrow(datas)
R> nl = nrow(datal)
R> dataH = data.frame(datas,
+        int_re = JointmodelCD4$summary.random$U12$mean[(nl+1):(nl+ns)],
+        slope_re = JointmodelCD4$summary.random$U22$mean[(nl+1):(nl+ns)])
R> dataL1 = data.frame(datal,
+        fitted_l = JointmodelCD4$summary.fitted.values$mean[1:nl],
+        random_l = JointmodelCD4$summary.random$U11$mean[1:nl],
+        randoms_l = JointmodelCD4$summary.random$U21$mean[1:nl])
\end{lstlisting}

\begin{figure}[H]
\centering
\includegraphics[width=16cm,height=16cm]{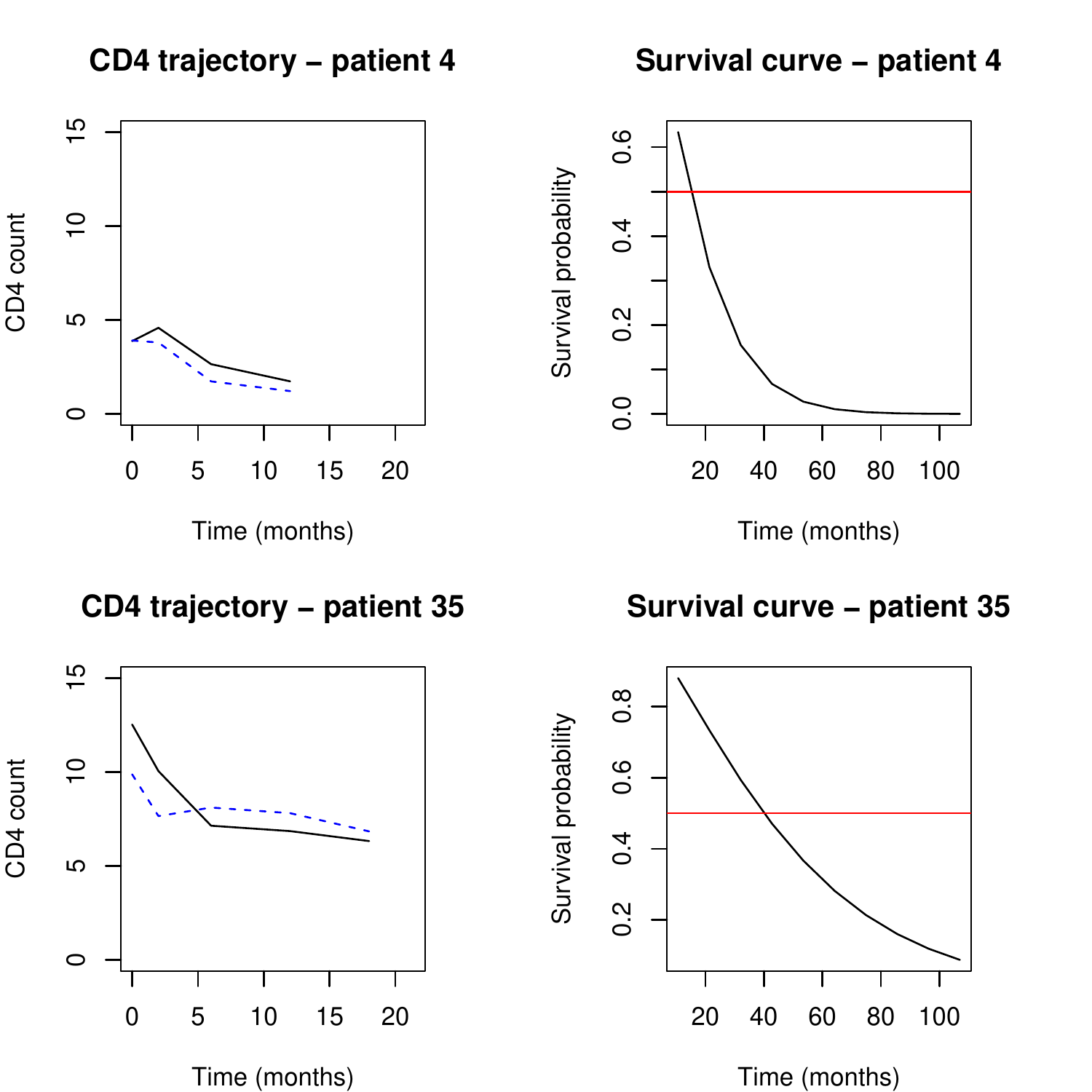}
\caption{Patient-specific plots}
\label{figaids}
\end{figure}

For illustration we produce the patient-specific CD4 trajectories and survival curves for two patients, one with AIDS infection at entry (patient 4) and one without (patient 35) in Figure \ref{figaids}.

\begin{lstlisting}

R> patients = c(4,35)
R> par(mfrow = c(2,2))
R> par(mar = c(4,4,4,4))
R> for (patientnr in patients){
+    dataHi = dataH[dataH$patient==patientnr,]
+    lambda = exp(-1.171 + 1.274 * (as.numeric(as.factor(dataHi$prevOI)) 
+              - 1) - 1.053 * dataHi$int_re)
+    alpha=1.282
+    plot(datal$obstime[datal$patient==patientnr] * 21.4,
+       datal$CD4[datal$patient==patientnr],ylab = "CD4 count",
+       xlab = "Time (months)",type = "l",xlim = c(0,21.4),
+       ylim = c(0,15),main = paste("CD4 trajectory - patient",patientnr))
+    lines(dataL1$obstime[dataL1$patient==patientnr] * 21.4,
+       (dataL1$fitted_l[dataL1$patient==patientnr]
+        + dataL1$random_l[dataL1$patient==patientnr]
+        + dataL1$randoms_l[dataL1$patient==patientnr]),
+        col="blue",lty=2)
+    plot(seq(0.1,1,0.1) * 21.4 * 5, exp(-((seq(0.1,1,0.1) * 5) ^ alpha) 
+       * (lambda + 0.993 * dataHi$slope_re * 5 * seq(0.1,1,0.1))),
+       type = "l",ylab = "Survival probability",xlab = "Time (months)",
+       main = paste("Survival curve - patient",patientnr))
+    abline(h = 0.5, col = "red")
+  }

\end{lstlisting}

\textbf{Example 2. PSA levels and informative dropout}

We follow \citet{hu2003} and \citet{kim2017} to estimate the longitudinal trajectory by
correcting for the bias introduced by informative dropout. Since the
main objective of the analysis is to estimate the non-linear
longitudinal trajectories while correcting the bias introduced by
informative dropout, we will use the entire longitudinal predictor as
the shared random effect i.e.\ $f(\eta^L_i(s))=\nu\eta^L_i(s)$. To model the non-linear trajectory we use a random walk order two component over time $\alpha(t)$. The model is thus:
\begin{eqnarray*}
        \eta^{L,J}_l(t)&=&\beta_0^L+\alpha(t)+\beta_1^L PSA_{base}\\
        \eta^{S,J}_i(s)&=&\beta_0^S+\nu \eta^{L,J}_i(s),
\end{eqnarray*}
where we assume a Weibull model for the dropout process.
Again, we preprocess the original data in the {JointModel} package. The resulting data set is available in the {INLA} package as {"exampledata/psa/jointdataPSA.rds"}.

\begin{figure}[H]
	\centering
\includegraphics[width=12cm]{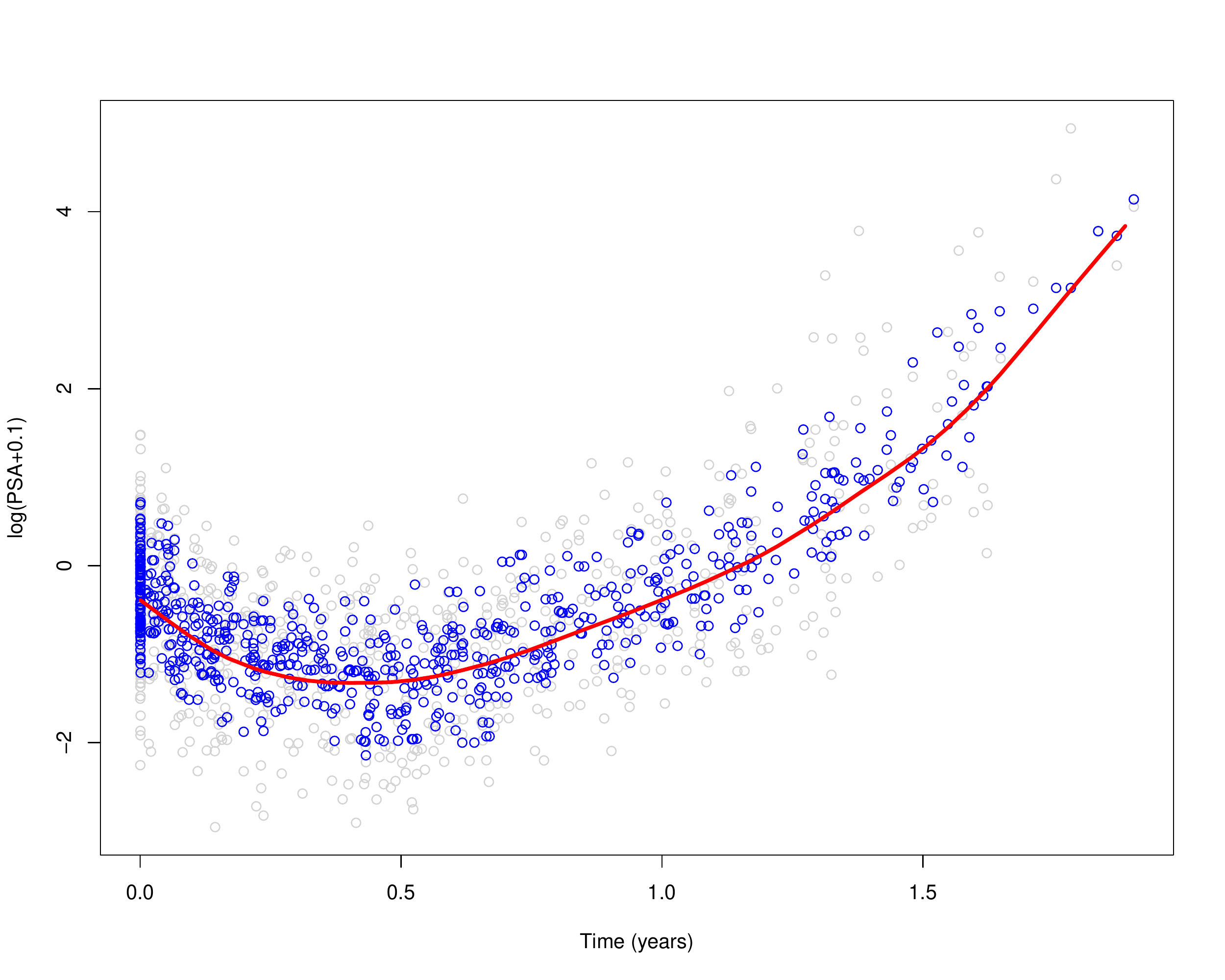}
\caption{Average PSA trajectory (red line) and estimated PSA levels (blue dots)}
\label{psaplot}
\end{figure}

\begin{lstlisting}
R> library("JointModel")
R> inla.setOption(short.summary = TRUE)
R> data1 = prostate
R> data2 = dropout
R> ng = nrow(data1)
R> ns = nrow(data2)
R> data1 = data1[order(data1$VisitTime),] 
R> joint.dataPSA = readRDS(system.file("exampledata/psa/jointdataPSA.rds",
+  package="INLA"))
R> Jointmodelres1 = inla(Y ~ -1 + mu + f(inla.group(V1, n = 50),model = 
+                  "rw2", scale.model = TRUE, hyper = list(prec = list
+                  (prior = "pc.prec", param = c(1, 0.01)))) + b13.PSAbase
+                  + f(u, w, model="iid", hyper = list(prec = list(initial
+                  = -6, fixed=TRUE))) + f(b.eta, copy="u", hyper = 
+                  list(beta = list(fixed = FALSE))), family = 
+                  c("gaussian","gaussian","weibullsurv"), data = 
+                  joint.dataPSA, verbose=TRUE, control.compute=list
+                  (dic = TRUE,config = TRUE), control.family = 
+                  list(list(),
+                    list(hyper = list(prec = list(initial = 10, 
+                    fixed = TRUE))), list()))
R> summary(Jointmodelres1)


Fixed effects:
              mean    sd 0.025quant 0.5quant 0.975quant   mode kld
mu1          0.083 0.038      0.008    0.083      0.157  0.083   0
mu2         -0.987 0.185     -1.366   -0.982     -0.640 -0.970   0
b13.PSAbase  0.421 0.026      0.370    0.421      0.472  0.421   0

Model hyperparameters:
                                         mean    sd 0.025quant
Precision for the Gaussian observations 2.084 0.112      1.870
alpha parameter for weibullsurv[3]      0.773 0.103      0.584
Precision for inla.group(V1, n = 50)    5.663 3.291      1.532
Beta for b.eta                          1.164 0.246      0.681
                                        0.5quant 0.975quant  mode
Precision for the Gaussian observations     2.08      2.313 2.078
alpha parameter for weibullsurv[3]          0.77      0.988 0.765
Precision for inla.group(V1, n = 50)        4.95     14.038 3.630
Beta for b.eta                              1.16      1.650 1.161

Deviance Information Criterion (DIC) ...............: -3750.68
Deviance Information Criterion (DIC, saturated) ....: 1184.64
Effective number of parameters .....................: 183.57

\end{lstlisting}
The association parameter $\nu$ is significant and it is thus clear that the joint model approach is necessary for this data set. The shape parameter for the Weibull model is estimated at $0.7863$ which is different from 1 and thus the exponential model would not suffice. The estimated non-linear longitudinal average trajectory is illustrated in Figure \ref{psaplot}.

\begin{lstlisting}
R> par(mfrow = c(1,1))
R> plot(data1$VisitTime,data1$logPSA.postRT,xlab = "Time (years)",
+       ylab = "log(PSA+0.1)",col = "lightgrey") 
R> points(data1$VisitTime,Jointmodelres1$summary.fitted.values[1:ng,1],
+         col = 'blue',lwd = 1)
R> lines(Jointmodelres1$summary.random$`inla.group(V1, n = 50)`[1:50,1],
+        Jointmodelres1$summary.random$`inla.group(V1, n = 50)`[1:50,2],
+        col = "red",lwd = 3)
\end{lstlisting}

\section{Non-separable space-time models}\label{stmodels}
The {INLA} package has been very successful in space and space-time modeling by representing spatial models with sparse matrices using the stochastic partial differential equations (SPDE) approach (\citet{lindgren2011explicit}, \citet{bakka2018spatial}, \citet{krainski2018advanced}).
Space-time models are usually constructed as Kronecker products, resulting in separable models, where the space-time covariance function is a product of a spatial and a temporal covariance function.
In {INLA} this is coded using the {group} and {control.group} arguments of the function {f}.

Instead of constructing a space-time model as an interaction between a spatial and a temporal model, \citet{bakka2019class} are developing a class of space-time models 
directly from the principles of diffusion processes in space-time.
The basic building block is a Mat\' ern model in space, which is smoothed by a space-time diffusion process. 
The spatial Mat\' ern model is a natural starting point due to its wide use in spatial modeling in general, and in {INLA} in particular.
Define the spatial differential operator
$$L =  \left( \gamma_s^2 - \Delta \right), $$
where $\gamma_s^2$ is a constant, and $\Delta = d^2/dx^2 + d^2/dy^2$ is the Laplacian.
The space-time diffusion process is governed by the differential operator
$$ \left( \gamma_t \frac{d}{dt} + L \right), $$
known as a reaction-diffusion operator in physics, and used in many physical models.
This operator is used in systems where mass 
(which can represent mass, energy, individuals, disease counts, or other characteristics) 
changes in time due to net transport from high value regions to low value regions (e.g.\ temperature equalises over time).

There is a rich literature on space-time 
models with non-separable covariance functions, 
see e.g.\ \citet{gneiting2002nonseparable}, \citet{stein2013class}, \citet{rodrigues2010class}, and references therein.
The literature focuses on constructing a model with a reasonable covariance function, 
and developing computationally efficient method for inference.
However, as far as we are aware, there are no flexible software implementations ready for use with these models.

In this section we discuss an implementation of the new models using the computational methods in {INLA}, by writing {R} code for the inference explicitly instead of using the {inla} function call.
This is both to reduce the computational time for this case study 
and to be transparent with all the details.
We recommend the reader to consider the sparsity structure of the 
matrices we present to see how well this approach fits with 
the research on parallel computations presented 
in Section \ref{parallel}.

\begin{lstlisting}
R> library("fields")
R> library("colorspace")
R> set.seed(2019)
\end{lstlisting}

We use simple temporal and spatial meshes as follows.
The spatial mesh can be plotted by {plot(mesh)} 
and is described in \citet{krainski2018advanced}.

\begin{lstlisting}
R> t.max = 8
R> mesh.time = inla.mesh.1d(1:t.max)
R> fake.locations = matrix(c(0,0,10,10, 0, 10, 10, 0), nrow = 4, byrow = T)
R> mesh.space = inla.mesh.2d(loc = fake.locations, max.edge=c(1.5, 2))
\end{lstlisting}

We use $\alpha_t=1, \alpha_s=2, \alpha_\epsilon=1$ in the model in \citet{bakka2019class}, and get 
the SPDE
$$L^{1/2} \left( \gamma_t \frac{d}{dt} + L \right) \gamma_\epsilon u(s, t) = \mathcal W (s,t), $$
where the $\gamma$'s are hyper-parameters, 
and $\mathcal W$ is a white noise process.

Before we can define the separable and the non-separable models, we need to decide the hyper-parameters for our two space-time models.
We use the empirical range for space and time
from \citet{lindgren2011explicit}.
The hyper-parameters for the separable model are as follows.

\begin{lstlisting}
R> range.time = 20 
R> range.space = 6
R> sigma.u = 1
\end{lstlisting}

We select hyper-parameters ($\gamma$'s) for the non-separable model to have a similar interpretation to
the interpretation of the parameters of the separable model \citep{bakka2019class}.

\begin{lstlisting}
R> gt = 2.23
R> gs2 = 8/range.space^2
R> ge2 = 0.0805

\end{lstlisting}

We use the finite element method (FEM) from \citet{lindgren2011explicit}, adopted by \citet{bakka2019class} to the following $M$-notation.
We note that the temporal model is first order Markov, and that
higher order Markov structure is used for models with a higher smoothness in time \citep{bakka2019class}.

\begin{lstlisting}
R> sfe = inla.mesh.fem(mesh.space, order = 4)
R> tfe = inla.mesh.fem(mesh.time, order = 2)
R> M0 = tfe$c0
R> stopifnot(abs(M0[1,1] - 0.5*M0[2,2])<1e-3)
R> N.t = nrow(M0)
R> M1 = sparseMatrix(i=c(1,N.t), j=c(1,N.t), x=0.5)
R> M2 = tfe$g1

\end{lstlisting}

The {sfe} and {tfe} objects contain the finite element matrices we need to build a solution to the SPDE.
Conditional on the chosen hyper-parameters, we define the precision matrices ($Q$) for the separable and the non-separable models.

\begin{lstlisting}
R> kappa = 2/range.time
R> Q.M = kappa^2*M0 + 2*kappa*M1 + M2
R> Q.M = Q.M/2/kappa
R> Q.space.alpha2 = gs2^2*sfe$c0 + 2*gs2*sfe$g1 + sfe$g2
R> Q.space.alpha2 = Q.space.alpha2/(4*pi*gs2)
R> Q.separ = kronecker(Q.M, Q.space.alpha2)
R> Q.nonsep = (kronecker(gt^2*M2, gs2*sfe$c0 + sfe$g1) + 
+              kronecker(M0, gs2^3*sfe$c0 + gs2^2*sfe$g1 +
+                     gs2*sfe$g2 + sfe$g3) +
+              kronecker(2*gt*M1, gs2^2*sfe$c0 +
+                     2*gs2*sfe$g1 + sfe$g2 )) * ge2

\end{lstlisting}

We can study the prior marginal variance and covariance structure as follows. 
Importantly, we note that the marginal variance is close to 1 for all models.

\begin{lstlisting}
R> print(diag(inla.qinv(Q.M)))
R> print(solve(Q.M)[1, ])
R> print(summary(diag(inla.qinv(Q.separ))))
R> print(summary(diag(inla.qinv(Q.nonsep))))
\end{lstlisting}

We simulate a Mat\'ern field for $t=1$. 
This can be done through the separable or the non-separable model, since they are both Mat\' ern marginally for $t=1$.
We use the {seed} and {num.threads=1} arguments to get reproducible simulations.
Further, we add a small noise to the observations to give a more realistic inference problem.
The dataframe {df} is represented for all of space and time, but we replace the observations by NA after year 1.

\begin{lstlisting}

R> u = inla.qsample(n=1, Q.nonsep, seed = 2019, num.threads=1)[, 1]
R> N.st = nrow(Q.separ)
R> sig.eps = 0.01
R> noise = rnorm(N.st, 0, 1) * sig.eps
R> df = data.frame(y=u+noise, st = 1:N.st)
R> df$y[-(1:mesh.space$n)] = NA
\end{lstlisting}

We follow the book \citet{rue2005} and compute posterior precision matrices and means, by conditioning on $y$.
Here, {Qeps} is the precision matrix for the observation noise, 
and {A.observe} is the matrix projecting from the latent field to the observation locations.

\begin{lstlisting}
R> Qeps = Diagonal(n=mesh.space$n)
R> A.observe = sparseMatrix(i=1:mesh.space$n, j=1:mesh.space$n, 
+                           dims = c(mesh.space$n, N.st))
R> ## Posterior/conditional precision matrix
R> post.Q = function(sig.eps=0.01, sig.v=1, Q.model) {
+    Q = sig.eps^-2*t(A.observe)%*%Qeps%*%A.observe+Q.model
+    return(Q)
+  }
R> ## Posterior mean (point estimate)
R> post.mu = function (sig.eps=0.01, sig.v=1, Q.model) {
+    a = df$y[1:mesh.space$n]
+    b = sig.eps^-2 * t(A.observe) %*% Qeps %*% a
+    res = inla.qsolve(post.Q(sig.eps, sig.v, Q.model = Q.model), b)
+    return(res)
+  } 
R> mu.post.separ = post.mu(Q.model = Q.separ)
R> mu.post.nonsep = post.mu(Q.model = Q.nonsep)

\end{lstlisting}

For convenience, we set up a local function for plotting, designed for our example.
This is developed from the code in \citet{krainski2018advanced}.

\begin{lstlisting}
R> local.plot.field = function(field, mesh, time=1, ...){
+    field = field[1:mesh$n + (time-1)*mesh$n]
+    proj = inla.mesh.projector(mesh, dims=c(200, 200))
+    field.proj = inla.mesh.project(proj, field)
+    image.plot(list(x = proj$x, y=proj$y, z = field.proj), 
+               col = diverging_hcl(63), ...)  
+  }

\end{lstlisting}

\begin{figure}[H]
\centering
\includegraphics[width=13cm]{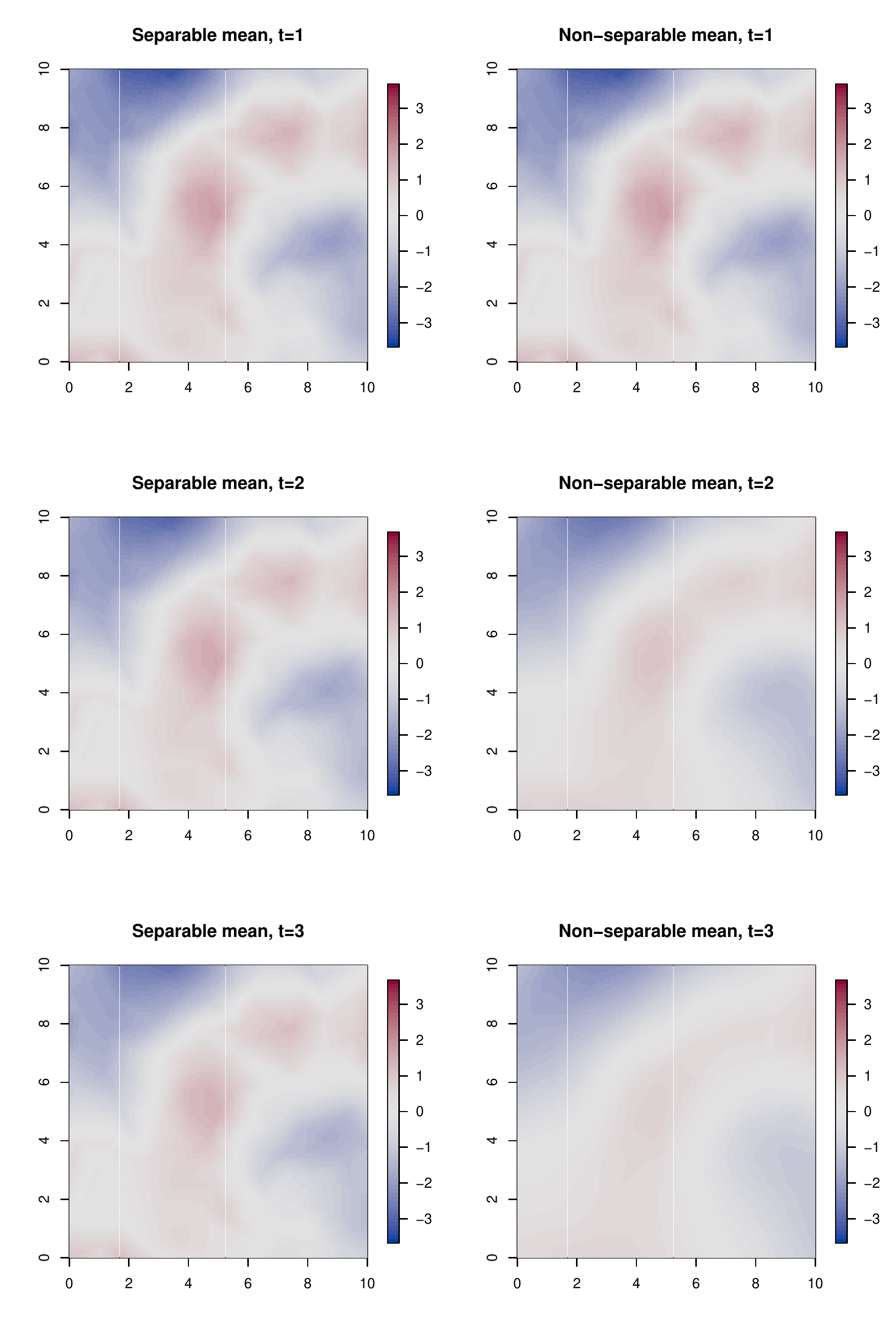}
\caption{Posterior mean estimates from the separable and non-separable models.}
\label{fignonsepmu}
\end{figure}

We plot the point predictions (posterior mean) in space-time, in Figure \ref{fignonsepmu}.
Note that in year 1 the field is conditioned on data on nearby locations, hence the separable and the non-separable models give very similar results.
Year 2 to 6, however, 
represent forecasts based on the data observed in year 1.
The plots shown here are for the first three years, the for loop can be extended to show all six.
In the figure we see a clear difference between the separable and the non-separable models.
The separable model forecasts a simple decay of the current observations, while the non-separable model results in smoother forecasts.
We argue that the non-separable forecast is more appropriate in most applied situations.
When forecasting e.g.\ the temperature in a location in the future, 
the model should use not just the temperature in the same location today,
but also use the temperature in nearby locations, 
resulting in a smoother forecast.
One classical example of this is hot water poured into cold water; we expect the two temperatures to regress to the mean by mixing and smoothing out differences.

\begin{lstlisting}
R> par(mfrow = c(3, 2))
R> zlim2 = c(-1, 1)*max(abs(c(mu.post.separ, mu.post.nonsep)))
R> for (tp in 1:3) {
+    local.plot.field(mu.post.separ, mesh.space, time = tp, 
+                     main = paste0("Separable mean, t=", tp),
+                     xlim=c(0, 10), ylim=c(0, 10), zlim=zlim2)
+    local.plot.field(mu.post.nonsep, mesh.space, time = tp, 
+                     main = paste0("Non-separable mean, t=", tp),
+                     xlim=c(0, 10), ylim=c(0, 10), zlim=zlim2)
+  }
\end{lstlisting}

\begin{figure}[H]
\centering
\includegraphics[width=13cm]{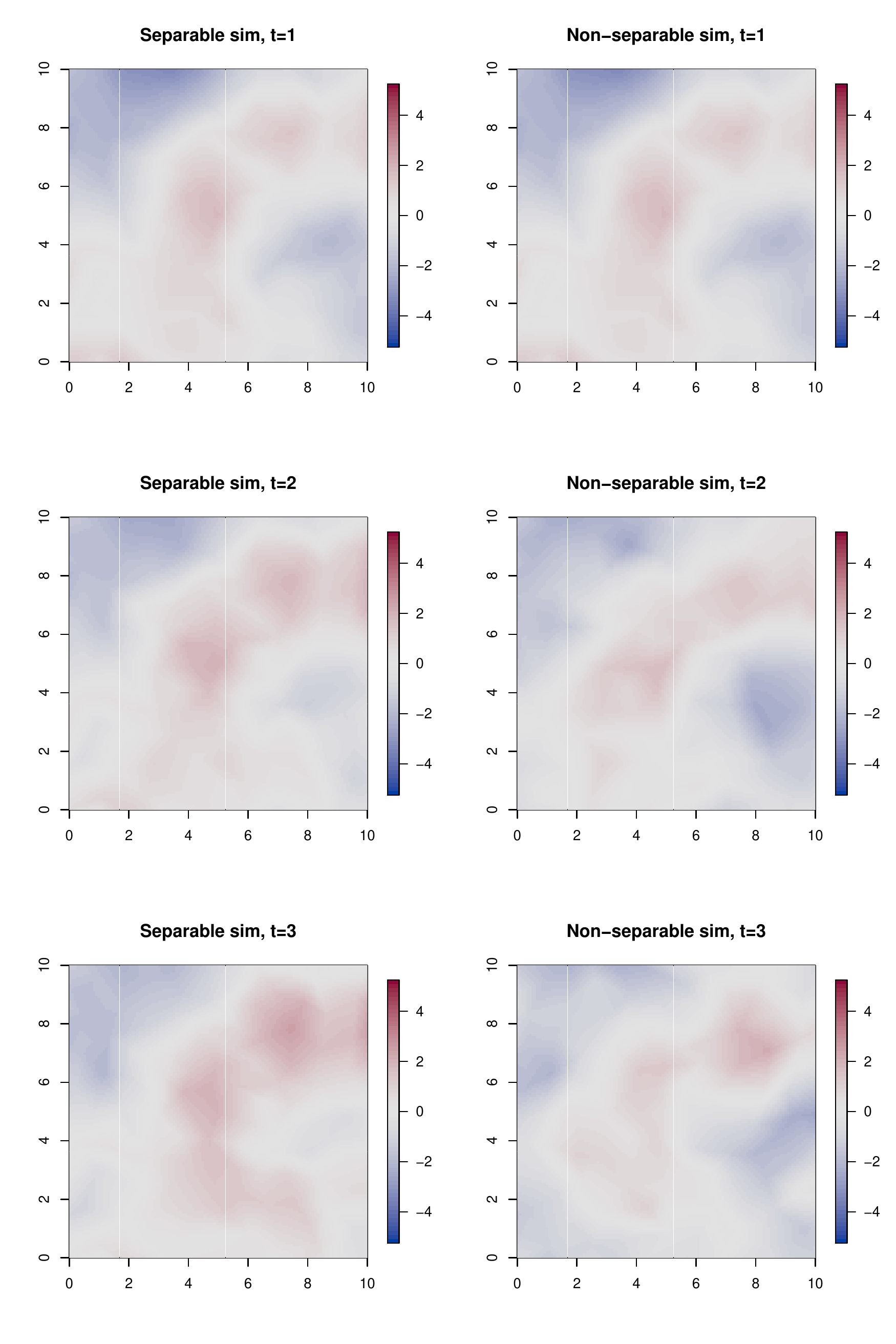}
\caption{Posterior simulations from the separable and non-separable models.}
\label{fignonsepsim}
\end{figure}

Finally, we show how to simulate from the posterior, in Figure \ref{fignonsepsim}.
As before, the first year is very similar, because we conditioned on data here, while year 2 and 3 show different simulations into the future.

\begin{lstlisting}
R> post.sim.separ = inla.qsample(1, Q=post.Q(Q.model = Q.separ), 
+                                reordering = "identity", seed = 1,
+                                num.threads=1)
R> post.sim.separ = drop(post.sim.separ + mu.post.separ)
R> post.sim.nonsep = inla.qsample(1, Q=post.Q(Q.model = Q.nonsep), 
+                                reordering = "identity", seed = 1,
+                                num.threads=1)
R> post.sim.nonsep = drop(post.sim.nonsep + mu.post.nonsep)
R> zlim1 = c(-1, 1)*max(abs(c(post.sim.separ, post.sim.nonsep)))
R> par(mfrow = c(3, 2))
R> for (tp in c(1,2,3)) {
+    local.plot.field(post.sim.separ, mesh.space, time = tp, 
+                     main = paste0("Separable sim, t=", tp),
+                     xlim=c(0, 10), ylim=c(0, 10), zlim=zlim1)
+    local.plot.field(post.sim.nonsep, mesh.space, time = tp, 
+                     main = paste0("Non-separable sim, t=", tp),
+                     xlim=c(0, 10), ylim=c(0, 10), zlim=zlim1)
+  }

\end{lstlisting}

In this code we used the option {reordering="identity"} in the {inla.qsample} function. 
The purpose of this is to use the same random noise, and the same reordering, to get a close comparison between the simulations.
In general, we recommend to use {inla.qsample} with a seed to get deterministic and reproducible behaviour, but to use the default reordering scheme to speed up computations.

\section[High performance and parallel computing]{High performance and parallel computing with the {INLA} package}\label{parallel}
The widespread acceptance of the INLA-approach and the {RINLA} software manifested as the {INLA} package,
was not foreseen when INLA was originally developed: hence, the {INLA}
package has continuously evolved from research code started
more than 15 years ago, adopting designs made for single-core
execution in mind. Today, there is a growing demand for analysing much
larger models: typically, either a large amount of observations and/or
a large number of latent variables (read space-time models, for
simplicity). And we have already started to provide better support for
the increasingly larger statistical models of today running on
computational platforms of tomorrow (typically multicore or manycore
and possibly hardware accelerated).

At the core of the INLA algorithm, is numerical linear algebra for
large sparse matrices. The tasks that is required, are for a symmetric
positive definite matrix $\mm{Q}$ of dimension $n$, the ability to
repeatedly compute
\begin{itemize}
\item the Cholesky factorization $\mm{Q} = \mm{L} \mm{L}^{T}$, where
    $\mm{L}$ is a lower triangular matrix,
\item solve linear systems like $\mm{L}\mm{x} = \mm{b}$,
    $\mm{L}^{T}\mm{x} = \mm{b}$, $\mm{L}\mm{L}^{T}\mm{x} = \mm{b}$,
    and
\item compute selected elements of the inverse of $\mm{Q}$,
    $(\mm{Q}^{-1})_{ij}$, for all $ij$ where $Q_{ij}$ is non-zero.
\end{itemize}
Additionally, we need also $\log|\mm{Q}|$, but since the Cholesky
factor is available, it is simply $\sum_i 2\log L_{ii}$. During the
whole INLA algorithm, the non-zero pattern of $\mm{Q}$ is the same,
which simplifies some of the initial procedures, like finding a good
reordering scheme.

For smaller $n$, like $n \sim 10^{4}$ to $10^{5}$ for a spatial model,
the serial algorithms for these tasks will run fine, as we have
parallelized (using OpenMP) on a higher level like factorizing several
matrices at once. For larger $n \sim 10^{5}$ to $10^{6}$, this
approach is no longer practical. Also, the type of model considered
plays a role here; space-time models is ${\mathcal O}(\sqrt{n})$ more
costly, and require more memory, than a spatial one, hence dimension
where the serial sparse matrix algorithms is no longer practical, will
be less.

The need for parallel numerical methods for large sparse matrices on
shared-memory and distributed-memory multiprocessors, have been
evident for quite some time. While there is a vast literature on the
development of efficient algorithms for the direct solution of sparse
linear systems of equations, only a few software package are
available, such as, e.g., MUMPS~(\cite{MUMPS:1,MUMPS:2}),
WSMP~(\cite{Gupta:2002:RAD}),
SuperLU~(\cite{Li:2005:OSA:1089014.1089017}),
CHOLMOD~(\cite{Davis:2006:DMS:1196434}). Neither of these libraries
provide parallel algorithms for all our required matrix operations
listed above, as they do not have a parallel implementation of the
algorithm to compute selected elements of the inverse. (CHOLMOD
support a serial version of this algorithm.) How to efficiently
compute selected elements of the inverse of a sparse matrix, have been
known for a quite some time \citep{pro20,art358}, but a parallel
version of this algorithm was not available in a main sparse matrix
library before the work of \cite{VERBOSIO201799} was made available in
the PARDISO library \citep{SCHENK2004475,
    10.1007/978-3-642-40047-6_54,doi:10.1137/130908737}. According to
\cite{Gould:2007:NES:1236463.1236465}, {PARDISO} one of the best
performing parallel libraries for numerical computations for large
sparse matrices.

A collaboration between
{PARDISO}~\footnote{\texttt{www.pardiso-project.org} Some may be aware
    of a former version of {PARDISO} which has been integrated into the
    Intel Math Kernel Library (MKL) a library of optimized math
    routines for science, engineering, and financial applications.}
and {INLA} project was initiated early 2018, ending up with a special
version of the {PARDISO} library for {INLA} which was integrated into
{INLA} and released in May 2018. With this new tool, we are now able
to run successfully statistical models with $n$ in the millions on
KAUST computational servers. The paralellisation strategy, that
currently is supported using argument
{control.compute = list(openmp.strategy = "pardiso.parallel")},
is to do one matrix at the time using a parallel algorithm to
factorize, solve and compute selected entries of the inverse. The
future plans for this collaboration, includes improvement of the
integration with the INLA algorithm including nested parallelism, and
also to extend the {PARDISO} interface so we can make use of more
efficiently computing capabilities exploiting the parallel computing
support in {PARDISO} to enable parallel distributed and accelerated
execution of the main numerical tasks required in the INLA algorithm.

To illustrate the abilities of the {PARDISO} library to work with huge
matrices, we ran a series of tests our computational server, with
512Gb of RAM, 2 sockets with 16 cores per socket, and with Intel(R)
Xeon(R) Gold 6130 CPUs @2.10GHz. The test matrix is constructed to be
very challenging, mimicking a large space-time model with the same
non-zero structure as the 3-dimensional Laplace equation on a
$n\times n\times n$ cube (which is the worst configuration).
Additionally, we added 25 dense rows/columns to mimic the presence of
fixed effects in the model. For the $(n^{3}+25) \times (n^{3}+25)$
sparse matrix, have about 56 neighbors for each node. The storage
required is about $0.22$Gb for $n=100$ and $1.72$Gb for $n=200$, to
store its non-zero elements. Additionally, we need to store their
(relative) location within the matrix.

\begin{figure}[H]
\centering
\includegraphics[width=0.5\linewidth]{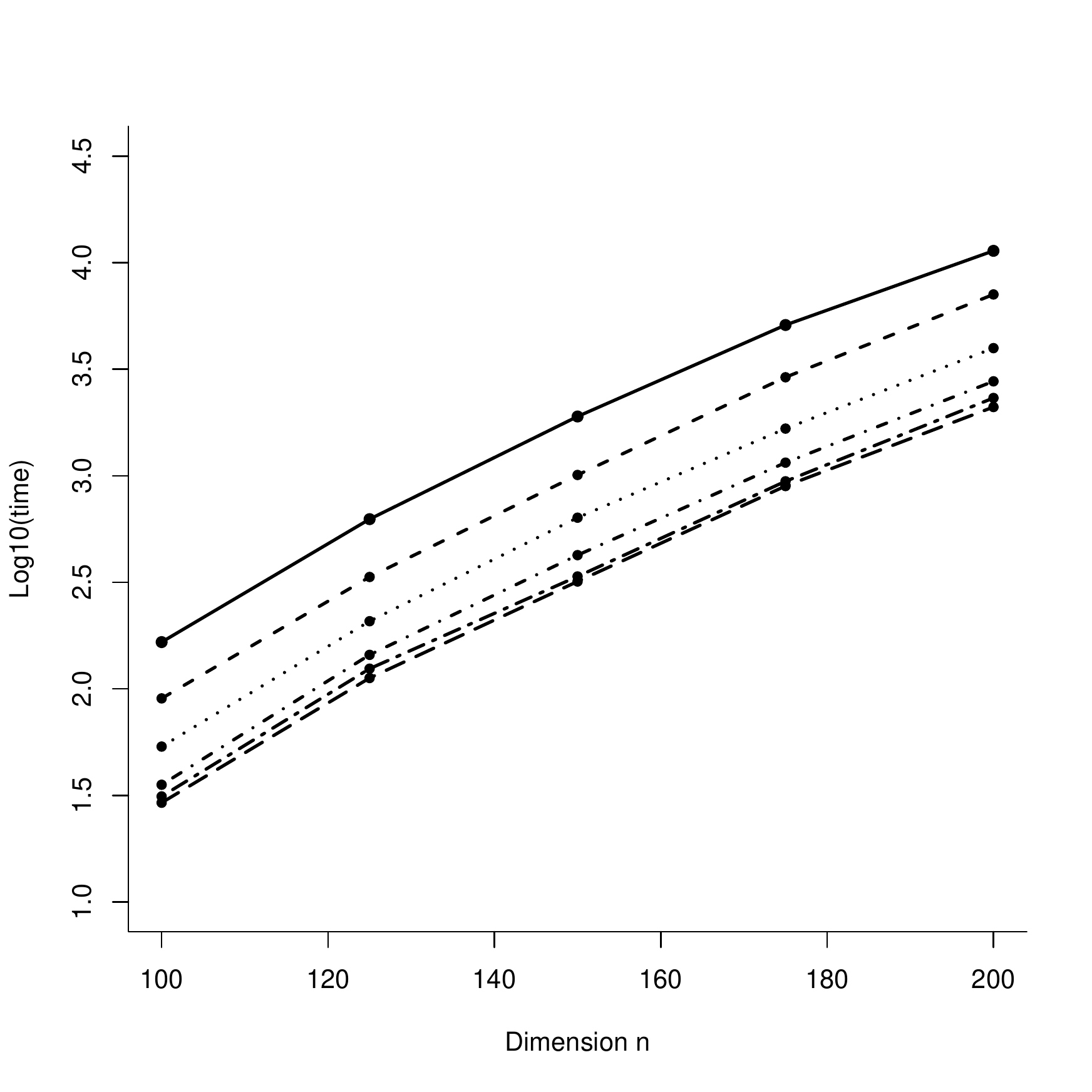}
\includegraphics[width=0.5\linewidth]{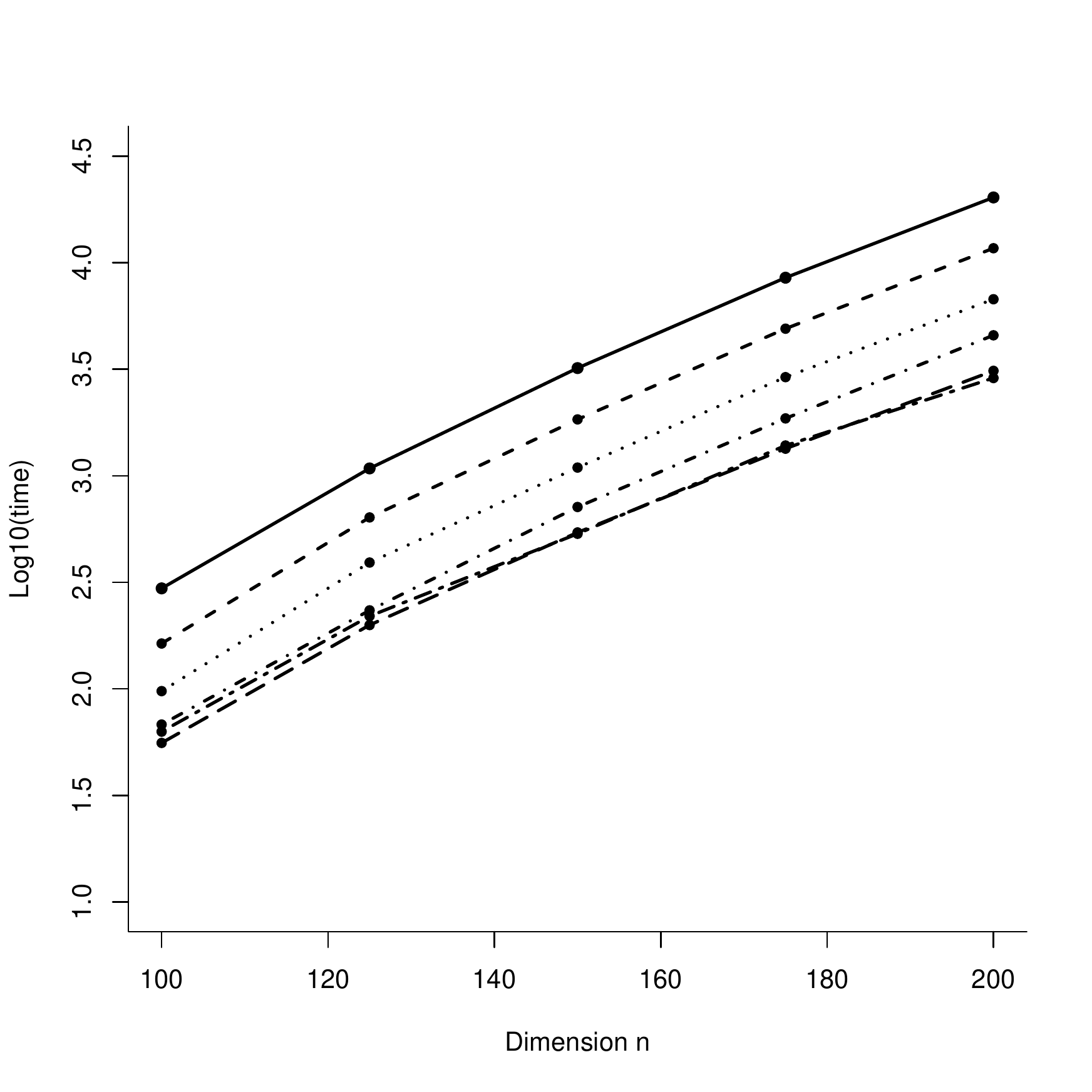} 
\caption{The running time doing Cholesky factorization (left) and computing the
        partial inverse for the 3D Laplace equation matrix with
        additional 25 dense row and columns. The dimension is
        $n^{3}+25$ with $n$ vary from 100 to 200. The number of cores
        are 1 (top), 2, 4, 8, 12, 16 and 32 (bottom).}
\label{parfig}
\end{figure}

Figure \ref{parfig} shows the results for $n=100, 120, 140, 160, 180$ and
$200$, using $nc=1, 2, 4, 8, 12, 16$ and $32$ cores, for doing
Cholesky factorization (left) and the partial inverse (right). The
results demonstrate a consistent behaviour for the running time both
with varying $n$ and $nc$. The computational cost reduces nicely from
$nc=1$ and 2 and to $4$, but then the speedup fades off. We do not
gain much going beyond 16 cores for this example, and the partial
inverse is somewhat more expensive to compute than the Cholesky
factorization. The results are very encouraging as it shows that
{PARDISO} can handle sparse matrices of this size and structure without
problems. The integration of {INLA} and {PARDISO} will be further
improved and we are currently working on this issue. 

\section{Discussion}

Bayesian modeling is ever present and still increasing in popularity
in applied fields of science. Initially, the inference was performed
using sampling-based methods like Gibbs sampling. These methods,
however, are often time-consuming and computationally inefficient.
From this impediment, approximate Bayesian inference approaches
sprouted. (One of) The most popular non-sampling based Bayesian
inference approach is the INLA methodology, facilitated through the
{INLA} package. Since the inception of INLA in 2009 through the
seminal paper \citep{rue2009}, the use of the INLA methodology has
been cited over $3000$ times. {INLA} is developed for the class of
latent Gaussian models, that contains most well-known statistical
models.

The success of {INLA} as a computational inferential framework for
Bayesian modeling is partly attributed to the continual development
and expansion of this package. As evident in this paper, relevant
statistical methodology is developed and implemented incessantly in
{INLA} as to provide scientists with a computational platform for
state-of-the-art Bayesian models.

The specific developments presented herein address some current
Bayesian modeling demands. In biomedical applications, the use of
joint models for survival and longitudinal data is imperative. The
efficacy of treatments as measured on multiple endpoints is a crucial
step in drug design, and necessitates the use of joint modeling of the
endpoints. In this paper, we presented the implementation of joint
models with one survival and one longitudinal endpoint. Future
developments in this field are under way and the need for a unique
interface for these joint models, based on the {INLA} architecture
is clear. The potential for further developments in this realm based
on {INLA} is encouraging. 
In the flavor of joint models, the extension
to spatial joint models, joint models with competing risks or
recurrent events and generalized multiple endpoint modeling are some
examples of models that could be implemented in {INLA} based on
the approach presented herein. Multistate models and competing risks
models are also of major interest in the biomedical field, and with
their implementation in {INLA} the extensions to spatial or
smoothing spline random effects would be trivial.

The innovative SPDE approach for space and space-time models as used in
{INLA} serves as a gateway for extensions in the field of space-time
modeling. The development of a class of non-separable space-time models is
motivated by current needs in the analysis of complex real space-time
data, and is based on physical diffusion processes. 
This extension is based on the definition of a particular SPDE
which is then solved using finite element methods, 
and contrasts to more common attempts at 
generalizing the covariance matrix or the spectrum.
This approach is unique to {INLA}
(within software for Bayesian modeling, as far as we know) and
ensures unequivocal computational efficiency, 
without additional approximations, 
compared to other
methods in the literature.

Based on the generalization to non-separable space-time models and the
increasing computational demand through big data, the ability of
{INLA} to perform in a high performance computing environment
necessitates the development of tools available in {INLA} that can
optimally facilitate the computational burden using high performance
computing architecture. To this end, we present the current and future
collaborative work on this front using the {PARDISO} library in
conjunction with {INLA}. This project promotes the use of
{INLA} to an even wider audience and ensures the applicability of
{INLA} for Bayesian inference in the future.

{INLA} equips the user with powerful Bayesian modeling tools that
are computationally efficient and relevant. The ongoing research and
development of {INLA} ensures congruence to state-of-the-art
statistical methodology and places the user at the vanguard of their
field.


\bibliographystyle{apalike}
\bibliography{BioJ,local-b,mybib,local-bibfile}

\begin{thebibliography}{}

\bibitem[Amestoy et~al., 2001]{MUMPS:1}
Amestoy, P.~R., Duff, I.~S., Koster, J., and L'Excellent, J.-Y. (2001).
\newblock A fully asynchronous multifrontal solver using distributed dynamic
  scheduling.
\newblock {\em SIAM Journal on Matrix Analysis and Applications}, 23(1):15--41.

\bibitem[Amestoy et~al., 2006]{MUMPS:2}
Amestoy, P.~R., Guermouche, A., L'Excellent, J.-Y., and Pralet, S. (2006).
\newblock Hybrid scheduling for the parallel solution of linear systems.
\newblock {\em Parallel Computing}, 32(2):136--156.

\bibitem[Andrinopoulou et~al., 2018]{andrinopoulou2018}
Andrinopoulou, E.-R., Eilers, P.~H., Takkenberg, J.~J., and Rizopoulos, D.
  (2018).
\newblock Improved dynamic predictions from joint models of longitudinal and
  survival data with time-varying effects using p-splines.
\newblock {\em Biometrics}, 74(2):685--693.

\bibitem[Bakka et~al., 2019]{bakka2019class}
Bakka, H., Krainski, E., Bolin, D., Rue, H., and Lindgren, F. (2019).
\newblock A class of interpretable and computationally efficient non-separable
  space-time models based on stochastic partial differential equations.
\newblock {\em arXiv preprint arXiv:xxx.xxx Work in progress (reference will be
  updated)}.

\bibitem[Bakka et~al., 2018]{bakka2018spatial}
Bakka, H., Rue, H., Fuglstad, G.-A., Riebler, A., Bolin, D., Illian, J.,
  Krainski, E., Simpson, D., and Lindgren, F. (2018).
\newblock Spatial modeling with {R-INLA}: A review.
\newblock {\em Wiley Interdisciplinary Reviews: Computational Statistics},
  10(6):e1443.

\bibitem[Braga et~al., 2018]{braga2018}
Braga, J., ter Braak, C.~J., Thuiller, W., and Dray, S. (2018).
\newblock Integrating spatial and phylogenetic information in the fourth-corner
  analysis to test trait--environment relationships.
\newblock {\em Ecology}, 99(12):2667--2674.

\bibitem[Cox, 1972]{cox1972}
Cox, D.~R. (1972).
\newblock Regression models and life-tables.
\newblock {\em Journal of the Royal Statistical Society: Series B
  (Methodological)}, 34(2):187--202.

\bibitem[Dalongeville et~al., 2018]{dalongeville2018}
Dalongeville, A., Benestan, L., Mouillot, D., Lobreaux, S., and Manel, S.
  (2018).
\newblock Combining six genome scan methods to detect candidate genes to
  salinity in the mediterranean striped red mullet (mullus surmuletus).
\newblock {\em BMC genomics}, 19(1):217.

\bibitem[Davis, 2006]{Davis:2006:DMS:1196434}
Davis, T.~A. (2006).
\newblock {\em Direct Methods for Sparse Linear Systems (Fundamentals of
  Algorithms 2)}.
\newblock Society for Industrial and Applied Mathematics, Philadelphia, PA,
  USA.

\bibitem[Erisman and Tinney, 1975]{art358}
Erisman, A.~M. and Tinney, W.~F. (1975).
\newblock On computing certain elements of the inverse of a sparse matrix.
\newblock {\em Communications of the {ACM}}, 18(3):177--179.

\bibitem[Gneiting, 2002]{gneiting2002nonseparable}
Gneiting, T. (2002).
\newblock Nonseparable, stationary covariance functions for space--time data.
\newblock {\em Journal of the American Statistical Association},
  97(458):590--600.

\bibitem[Gould et~al., 2007]{Gould:2007:NES:1236463.1236465}
Gould, N. I.~M., Scott, J.~A., and Hu, Y. (2007).
\newblock A numerical evaluation of sparse direct solvers for the solution of
  large sparse symmetric linear systems of equations.
\newblock {\em ACM Trans. Math. Softw.}, 33(2).

\bibitem[Graetz et~al., 2018]{graetz2018}
Graetz, N., Friedman, J., Osgood-Zimmerman, A., Burstein, R., Biehl, M.~H.,
  Shields, C., Mosser, J.~F., Casey, D.~C., Deshpande, A., Earl, L., et~al.
  (2018).
\newblock Mapping local variation in educational attainment across africa.
\newblock {\em Nature}, 555(7694):48.

\bibitem[Guo and Carlin, 2004]{guo2004}
Guo, X. and Carlin, B.~P. (2004).
\newblock Separate and joint modeling of longitudinal and event time data using
  standard computer packages.
\newblock {\em The American Statistician}, 58(1):16--24.

\bibitem[Gupta, 2002]{Gupta:2002:RAD}
Gupta, A. (2002).
\newblock Recent advances in direct methods for solving unsymmetric sparse
  systems of linear equations.
\newblock {\em {ACM} Transactions on Mathematical Software}, 28(3):301--324.

\bibitem[Henderson et~al., 2000]{henderson2000}
Henderson, R., Diggle, P., and Dobson, A. (2000).
\newblock Joint modelling of longitudinal measurements and event time data.
\newblock {\em Biostatistics}, 1(4):465--480.

\bibitem[Hu and Sale, 2003]{hu2003}
Hu, C. and Sale, M.~E. (2003).
\newblock A joint model for nonlinear longitudinal data with informative
  dropout.
\newblock {\em Journal of Pharmacokinetics and Pharmacodynamics},
  30(1):83--103.

\bibitem[Ibrahim et~al., 2010]{ibrahim2010}
Ibrahim, J.~G., Chu, H., and Chen, L.~M. (2010).
\newblock Basic concepts and methods for joint models of longitudinal and
  survival data.
\newblock {\em Journal of Clinical Oncology}, 28(16):2796.

\bibitem[Kim et~al., 2017]{kim2017}
Kim, S., Zeng, D., and Taylor, J.~M. (2017).
\newblock Joint partially linear model for longitudinal data with informative
  drop-outs.
\newblock {\em Biometrics}, 73(1):72--82.

\bibitem[Krainski et~al., 2019]{krainski2018advanced}
Krainski, E.~T., G{\'o}mez-Rubio, V., Bakka, H., Lenzi, A., Castro-Camilio, D.,
  Simpson, D., Lindgren, F., and Rue, H. (2019).
\newblock {\em Advanced Spatial Modeling with Stochastic Partial Differential
  Equations using {R} and {INLA}}.
\newblock New York: Chapman and Hall/CRC.
\newblock Github version www.r-inla.org/spde-book.

\bibitem[Kuzmin et~al., 2013]{10.1007/978-3-642-40047-6_54}
Kuzmin, A., Luisier, M., and Schenk, O. (2013).
\newblock Fast methods for computing selected elements of the green's function
  in massively parallel nanoelectronic device simulations.
\newblock In Wolf, F., Mohr, B., and an~Mey, D., editors, {\em Euro-Par 2013
  Parallel Processing}, pages 533--544, Berlin, Heidelberg. Springer Berlin
  Heidelberg.

\bibitem[Li, 2005]{Li:2005:OSA:1089014.1089017}
Li, X.~S. (2005).
\newblock An overview of {SuperLU}: Algorithms, implementation, and user
  interface.
\newblock {\em ACM Trans. Math. Softw.}, 31(3):302--325.

\bibitem[Lindgren et~al., 2011]{lindgren2011explicit}
Lindgren, F., Rue, H., and Lindstr{\"o}m, J. (2011).
\newblock An explicit link between {G}aussian fields and {G}aussian {M}arkov
  random fields: the stochastic partial differential equation approach.
\newblock {\em Journal of the Royal Statistical Society B}, 73(4):423--498.

\bibitem[Martino et~al., 2011]{martino2011}
Martino, S., Akerkar, R., and Rue, H. (2011).
\newblock Approximate {Bayesian} inference for survival models.
\newblock {\em Scandinavian Journal of Statistics}, 38(3):514--528.

\bibitem[Osgood-Zimmerman et~al., 2018]{osgood2018}
Osgood-Zimmerman, A., Millear, A.~I., Stubbs, R.~W., Shields, C., Pickering,
  B.~V., Earl, L., Graetz, N., Kinyoki, D.~K., Ray, S.~E., Bhatt, S., et~al.
  (2018).
\newblock Mapping child growth failure in africa between 2000 and 2015.
\newblock {\em Nature}, 555(7694):41.

\bibitem[Petra et~al., 2014]{doi:10.1137/130908737}
Petra, C., Schenk, O., Lubin, M., and Gäertner, K. (2014).
\newblock An augmented incomplete factorization approach for computing the
  schur complement in stochastic optimization.
\newblock {\em SIAM Journal on Scientific Computing}, 36(2):C139--C162.

\bibitem[Podschwit et~al., 2018]{podschwit2018}
Podschwit, H., Larkin, N., Steel, E., Cullen, A., and Alvarado, E. (2018).
\newblock Multi-model forecasts of very-large fire occurences during the end of
  the 21st century.
\newblock {\em Climate}, 6(4):100.

\bibitem[Quintero and Jetz, 2018]{quintero2018}
Quintero, I. and Jetz, W. (2018).
\newblock Global elevational diversity and diversification of birds.
\newblock {\em Nature}, 555(7695):246.

\bibitem[Ratcliffe et~al., 2004]{ratcliffe2004}
Ratcliffe, S.~J., Guo, W., and Ten~Have, T.~R. (2004).
\newblock Joint modeling of longitudinal and survival data via a common
  frailty.
\newblock {\em Biometrics}, 60(4):892--899.

\bibitem[Rodrigues and Diggle, 2010]{rodrigues2010class}
Rodrigues, A. and Diggle, P.~J. (2010).
\newblock A class of convolution-based models for spatio-temporal processes
  with non-separable covariance structure.
\newblock {\em Scandinavian Journal of Statistics}, 37(4):553--567.

\bibitem[Rodr{\'\i}guez~de Rivera et~al., 2018]{rodriguez2018}
Rodr{\'\i}guez~de Rivera, {\'O}., L{\'o}pez-Qu{\'\i}lez, A., and Blangiardo, M.
  (2018).
\newblock Assessing the spatial and spatio-temporal distribution of forest
  species via bayesian hierarchical modeling.
\newblock {\em Forests}, 9(9):573.

\bibitem[Rue and Held, 2005]{rue2005}
Rue, H. and Held, L. (2005).
\newblock {\em Gaussian {Markov} random fields: theory and applications}.
\newblock CRC press.

\bibitem[Rue et~al., 2009]{rue2009}
Rue, H., Martino, S., and Chopin, N. (2009).
\newblock Approximate {Bayesian} inference for latent {Gaussian} models by
  using integrated nested laplace approximations.
\newblock {\em Journal of the Royal Statistical Society B}, 71(2):319--392.

\bibitem[Schenk and G{\"a}rtner, 2004]{SCHENK2004475}
Schenk, O. and G{\"a}rtner, K. (2004).
\newblock Solving unsymmetric sparse systems of linear equations with
  {PARDISO}.
\newblock {\em Future Generation Computer Systems}, 20(3):475 -- 487.
\newblock \textbf{Cited$>$1200}.

\bibitem[Shaddick et~al., 2018]{shaddick2018}
Shaddick, G., Thomas, M.~L., Amini, H., Broday, D., Cohen, A., Frostad, J.,
  Green, A., Gumy, S., Liu, Y., Martin, R.~V., et~al. (2018).
\newblock Data integration for the assessment of population exposure to ambient
  air pollution for global burden of disease assessment.
\newblock {\em Environmental science \& technology}, 52(16):9069--9078.

\bibitem[Stein, 2013]{stein2013class}
Stein, M.~L. (2013).
\newblock On a class of space--time intrinsic random functions.
\newblock {\em Bernoulli}, 19(2):387--408.

\bibitem[Stuart-Smith et~al., 2018]{stuart2018}
Stuart-Smith, R.~D., Brown, C.~J., Ceccarelli, D.~M., and Edgar, G.~J. (2018).
\newblock Ecosystem restructuring along the great barrier reef following mass
  coral bleaching.
\newblock {\em Nature}, 560(7716):92.

\bibitem[Takahashi et~al., 1973]{pro20}
Takahashi, K., Fagan, J., and Chen, M.~S. (1973).
\newblock Formation of a sparse bus impedance matrix and its application to
  short circuit study.
\newblock In {\em 8th PICA Conference proceedings}, pages 63--69. IEEE Power
  Engineering Society.
\newblock Papers presented at the $1973$ Power Industry Computer Application
  Conference in Minneapolis, Minnesota.

\bibitem[Van~Niekerk et~al., 2019]{van2019}
Van~Niekerk, J., Bakka, H., and Rue, H. (2019).
\newblock Joint models as latent gaussian models-not reinventing the wheel.
\newblock {\em arXiv preprint arXiv:1901.09365}.

\bibitem[Verbosio et~al., 2017]{VERBOSIO201799}
Verbosio, F., Coninck, A.~D., Kourounis, D., and Schenk, O. (2017).
\newblock Enhancing the scalability of selected inversion factorization
  algorithms in genomic prediction.
\newblock {\em Journal of Computational Science}, 22:99 -- 108.

\bibitem[Wulfsohn and Tsiatis, 1997]{wulfsohn1997}
Wulfsohn, M.~S. and Tsiatis, A.~A. (1997).
\newblock A joint model for survival and longitudinal data measured with error.
\newblock {\em Biometrics}, pages 330--339.

\bibitem[Zhou et~al., 2008]{zhou2008}
Zhou, H., Lawson, A.~B., Hebert, J.~R., Slate, E.~H., and Hill, E.~G. (2008).
\newblock Joint spatial survival modeling for the age at diagnosis and the
  vital outcome of prostate cancer.
\newblock {\em Statistics in medicine}, 27(18):3612--3628.

\end{thebibliography}



\end{document}